# Unified Power System Analyses and Models using Equivalent Circuit Formulation

Amritanshu Pandey, Marko Jereminov, Xin Li, Gabriela Hug, Larry Pileggi
Dept. of Electrical and Computer Engineering
Carnegie Mellon University
Pittsburgh, PA

*Abstract* — In this paper we propose and demonstrate the potential for unifying models and algorithms for the steady state and transient simulation of single-phase and three-phase power systems. At present, disparate algorithms and models are used for the different analyses, which can lead to inconsistencies – such as the transient analysis as time approaches infinity not matching the steady state analysis of the same conditions. Using our equivalent circuit formulation of the power system, we propose a methodology for forming physics-based models that can facilitate transient, balanced power flow, and three-phase power flow in one simulation environment. The approach is demonstrated on a three-phase induction motor. Existing industry tools are used to validate the model and simulation results for the different analyses.

## I. INTRODUCTION

Present-day computer modeling and simulation techniques for frequency domain (steady state) and time domain power system analyses were first introduced several decades ago [9]. Both the frequency domain and time domain analyses evolved independently over time and used different modeling and simulation techniques to obtain system solutions [11]-[12]. The frequency domain analysis, commonly referred to as power flow analysis, yields the steady state solution of the system through solving the power balance equations with node voltages and its angles as unknown variables. The time domain analysis yields the dynamic response of the system using circuit based modified nodal method (MNA) to solve the system of equations with voltages and currents as unknown variables.

In general, the expectation should be for the final steady state of the transient response to match exactly with the balanced power flow solution or the three-phase power flow solution of the system. However, this is generally not the case. The nonuse of standardized models and simulation algorithms between the different power system analyses often leads to inconsistent and erroneous results. This is in contrast to circuit simulation that is used for electronic systems [7], wherein standardization of models and algorithms guarantees consistent results between steady state and transient analyses.

A most notable modeling difference is that loads and generators in power flow analysis are modeled using non-physics based real and reactive average power variables (PV/PQ models). These variables are time average magnitudes with phasor relationship and are, therefore, inherently incompatible with time domain analysis. Due to this, time domain analysis either uses physics based models or some form of approximation of the constant power models (e.g. constant impedance) to model loads and generators. This modeling discrepancy between the two analyses often yields inconsistent results.

In the past, the use of real and reactive power variables to model aggregated load and generation in power flow analysis was necessary due to lack of real synchronized measurement data for the power grid. However, the advent of phasor measurement units (PMUs) with time stamped voltage and current measurements allows for aggregated load characterization using real measurement data with voltage and current as unknown variables [13]. We have recently demonstrated a circuit-based formulation for the steady state analysis of power systems [1]-[3] that is based on the use of these physical voltage and current state variables. The formulation allows for amalgamation of aggregated load models and physics based detailed models that are incompatible with conventional power flow. Importantly, this can provide a first step toward standardization of steady state and transient models and unification of power system analyses.

In this paper, we describe the formulation and analyses that can provide consistency between steady state and transient solutions using unified physics-based models. A simple model of a three-phase squirrel cage induction motor is used for demonstration. The example system is modeled using our equivalent circuit formulation [1]-[3] and solved using a tree link (TLA) analysis formulation for the unknown branch voltages and currents. Our prototype tool, <u>S</u>imulation with <u>U</u>nified <u>G</u>rid <u>A</u>nalyses and <u>R</u>enewables (SUGAR) is used to run the transient and steady state analysis of the example system, independently. The results from SUGAR are compared against an industry tool to validate the consistent results between the two analyses.

## II. BACKGROUND

### A. Power Flow Analysis

An equivalent circuit formulation with voltage and current as state variables was previously introduced in [1] - [3] to perform frequency domain balanced and three-phase power flow analyses. It was demonstrated that the power grid components and bus models could be modelled as combinations of circuit elements, i.e. impedances, voltage, and current sources, without loss of generality. The non-linear complex variable based bus models (PQ and PV) were formulated using a novel split circuit approach as follows:

*1) Newton Raphson with Complex Variables and Split Circuit Approach*

To obtain the complete equivalent circuit of a power grid, each power system component, i.e. transmission line, PQ load, PV bus, etc. is translated to an equivalent circuit model as derived in [1] – [3]. However, some of these circuit elements are found to be nonlinear, which necessitates the use of a nonlinear solution method. Due to non-analyticity of conjugate based complex functions, the preferred Newton-Raphson (N-R) method cannot be applied directly since it involves taking a first-order Taylor expansion of the non-linear equations. A key insight in handling of this problem is to use the split circuit approach. Following the split circuit approach, the equivalent circuit of the power system is split into real and imaginary sub-circuits coupled by controlled sources. The resulting coupled sub-circuits are then described by real functions and, therefore, are differentiable, which allows N-R to be applied.

*2) Power Flow load and generator models*

In equivalent circuit formulation, both load and generator (PV and PQ) are modeled as non-linear voltage controlled current sources:

$$I_{RB} = \frac{P_B V_{RG} + Q_B V_{IG}}{V_{RB}^2 + V_{IB}^2} \quad (1)$$

$$I_{IB} = \frac{P_B V_{IG} - Q_B V_{RG}}{V_{RB}^2 + V_{IB}^2} \quad (2)$$

where subscript $B$ represents generator (G) or load (L).

It should be noted that for the generator model the reactive power $Q_G$ is unknown, so it is added as a variable. Further, an extra equation representing a constraint that keeps the generator bus voltage magnitude constant is added to keep the number of equations and variables consistent.

### B. Transient Analysis

Transient analysis of a power grid involves evaluating the equivalent circuit in time domain. The system of first-order nonlinear differential algebraic equations (DAEs) is solved as follows:

*1) Taylor's First Order Approximation and Trapezoidal Rule*

Non-linear transient analysis begins with linearizing the non-linear system of equations using Taylor's first-order approximation, just as it is done for the case of non-linear power flow analysis in [1] - [3]. For example, speed-voltage term: $f(\omega_r, I_{ds})$ of the induction machine with rotor speed ($\omega_r$) and direct-axis stator current ($I_{ds}$) as state variables is linearized as follows:

$$f^{k+1}(\omega_r, I_{ds}) = f^k(\omega_r, I_{ds}) + (\omega_r^{k+1} - \omega_r^k) f'_{\omega_r} \\ + (I_{ds}^{k+1} - I_{ds}^k) f'_{I_{ds}} \quad (3)$$

With this approximation, the linearized system of equations is representable by respective circuit elements (voltage and current sources, passive elements, etc.).

To convert the system of equations to a nonlinear algebraic set that can be solved via N-R, the time derivative terms are first converted to time difference terms using trapezoidal numerical integration. In doing so, all time derivative elements in the equivalent circuit are replaced by their companion models [4].

For example, consider the two-coil model shown in Figure 1. The voltage equation for the first coupled coil is as follows:

$$V_1 = L_{11} \frac{dI_1}{dt} + L_{12} \frac{dI_2}{dt} \quad (4)$$

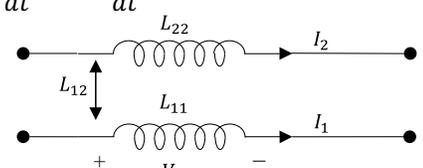

Figure 1: *Two coil example for trapezoidal rule illustration*

Using the trapezoidal rule, the differential equation (4) can be represented by the following difference equation:

$$V_{1(t+\Delta t)} = -V_{11}^L(t) - \frac{2L_{11}}{\Delta t} I_1(t) + \frac{2L_{11}}{\Delta t} I_{1(t+\Delta t)} - V_{12}^L(t) \\ - \frac{2L_{12}}{\Delta t} I_2(t) + \frac{2L_{12}}{\Delta t} I_{2(t+\Delta t)} \quad (5)$$

The difference equation (5) can be further rearranged and reduced to following form:

$$V_{1(t+\Delta t)} = -V_{11}^L(t) - V_{12}^L(t) - \frac{2L_{12}}{\Delta t} I_2(t) - \frac{2L_{12}}{\Delta t} I_1(t) \\ + R_{Eq} I_{1(t+\Delta t)} + \frac{2L_{11}}{\Delta t} I_{2(t+\Delta t)} \quad (6)$$

$$V_{1(t+\Delta t)} = -V_{1\_hist} + R_{Eq} I_{1(t+\Delta t)} + \frac{2L_{12}}{\Delta t} I_{2(t+\Delta t)} \quad (7)$$

The voltage at time $t + \Delta t$ across the coil 1 is approximated using (7) and is stamped into the equivalent circuit, as described in [4] and shown in Figure 2. The first term is only dependent on the historical magnitudes of the variables and can be represented by an independent voltage source. Similarly, a resistance can represent the second term and a current controlled voltage source can represent the third term.

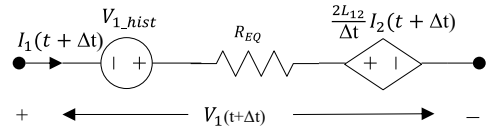

Figure 2: *Equivalent circuit for the two-coil example*

Once the resulting time-discretized equivalent circuit is linearized using the first two terms of the Taylor series approximation, just as it was done for the steady state solution, the circuit is solved iteratively for each time-point using the circuit-based models and formulation. Specifically, the N-R method is used at each time step to solve the non-linear equivalent circuit to an acceptable accuracy.

## III. UNIFYING POWER SYSTEM SIMULATION

Consistent results between the steady state and transient analysis is essential for precisely studying the behavior of the power system. Particularly, the steady state solution of the system should match exactly with the transient response of the system as time approaches infinity. To achieve this goal, our equivalent circuit formulation is extended to incorporate physics-based first-principle models for power system components and a compatible simulation algorithm. This will allow us to apply some of the same circuit simulation methods (e.g. SPICE [7] and its many derivatives [4]) that are used today to routinely simulate semiconductor circuits with millions of nonlinear transistors to ensure robust convergence to a realistic physics-based system solution. This equivalent circuit formulation further allows us to explore the use of nonlinear steady state formulations for circuits, such as harmonic balance [15], for capturing the frequency harmonics induced into a power grid by nonlinear components or unwanted disruptions.

### A. Physics based Models

Models form the basis for most power system studies [8]. Therefore, in order to model a power system accurately, we need to look beyond non-physics based conventional PQ and PV models that presently are used in power flow analysis. The use of physics-based models, with voltage and currents as state variables, models the system in its most realistic physical state and thus yields consistent and accurate results between the steady state and transient analyses. Furthermore, the use of natural system variables to model the power system will best emulate the measured system conditions.

### B. Tree Link for Transient Analysis

To formulate the circuit equations using the equivalent circuit components, a graph-theoretic tree-link (TLA) method is applied to solve for the circuit voltages and currents. TLA has already been shown to perform seamlessly for balanced power flow and three-phase power flow analysis [1]–[3], offering superior robustness over existing nodal methods (MNA) [14]. MNA is generally used for circuit simulation of electronic systems due to its simplicity and efficiency, however, TLA is known to provide superior numerical conditioning and the ability to accommodate both voltage and current state variables inherently.

For applications in three-phase power system analyses, the ability of TLA to naturally incorporate current state variables enables the handling of a large number of coupled inductors. Furthermore, the TLA formulation is capable of accommodating ideal switches (switching from zero impedance to zero conductance), which enables the power flow simulations to include components that are switched into and out of the grid. This capability is particularly helpful for efficient contingency analyses.

## IV. THREE-PHASE SQUIRREL CAGE INDUCTION MOTOR MODEL

A three-phase squirrel cage induction motor (IM) is used to demonstrate the unified analyses presented in this paper. The IM is a non-linear circuit. The set of equations that define the behavior of an IM are stiff due to system poles that are spread out in the complex plain. The mechanical sub-system of an IM has a very slow system response and has a pole that is very close to the imaginary axis of the complex plane. The electrical sub-system of an IM has a very fast response and has poles located far from the imaginary axis of the complex plane.

The IM models for the purposes of power flow analyses (frequency domain) differ significantly from those used in transient analysis. It is often observed that for the purposes of balanced and three-phase power flow analyses, operating rotor speed of the IM is assumed, thereby making the system of equations linear in nature. However, in case of transient analysis, the non-linear set of equations is solved to find the rotor speed. This discrepancy usually results in inconsistent results between the transient solution of the IM and the balanced three-phase power flow solution of the IM. The approach described above will use the same model for both analyses and yields consistent results between the steady-state analysis and the transient analysis.

### A. DQ Transformation

The flux generated by the three-phase IM in abc frame has time varying coefficients in its voltage terms due to the sinusoidal nature of the mutual inductance. This makes the analysis of three phase IM cumbersome in the abc reference frame. However, this undesirable feature can be eliminated by use of dq transformation. Dq transformation can be performed by choosing one of the three reference frames: i) synchronous reference frame; ii) stationary reference frame; and iii) rotating reference frame.

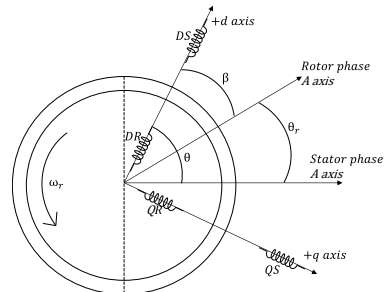

Figure 3: *Superimposition of dq-axis on 3-phase induction motor*

The final response of the IM is independent of the chosen reference frame. However, each of the reference frames has its own advantages and disadvantages depending on the problem that is being investigated [5]. For the purposes of this paper, we use the synchronously rotating reference frame.

The dq transformation matrix $P_\theta$ for the stator variable transformation is as follows:

$$[P_\theta] = \frac{2}{3}\begin{bmatrix} 0.5 & 0.5 & 0.5 \\ \cos(\theta) & \cos(\theta-\lambda) & \cos(\theta+\lambda) \\ \sin(\theta) & \sin(\theta-\lambda) & \sin(\theta+\lambda) \end{bmatrix} \quad (8)$$

and,

$$[F_{0dq}] = [P_\theta] \cdot [F_{abc}] \quad (9)$$

where function $F$ can represent either currents or voltages.

For rotor variable transformation, $\theta$ is replaced with $\beta$ in the equations above. For synchronous reference frame, the machine angle and speed variables are defined as follows:

$$\omega = p\theta = \omega_s \quad (10)$$
$$\beta = \theta - \theta_r = \theta_s - \theta_r \quad (11)$$

where $p$ is the differential operator. $\omega_s$ and $\omega_r$ are the synchronous and rotor speed of the motor, respectively, and $\theta_s$ and $\theta_r$ are the stator and rotor position, respectively.

B. *Motor Equations in Transient Domain*

The set of electrical equations that define the dynamic behavior of the IM are as follows [6]:

$$v_{ds} = R_s I_{ds} + p\psi_{ds} - \psi_{qs} p\theta \quad (12)$$
$$v_{qs} = R_s I_{qs} + p\psi_{qs} + \psi_{ds} p\theta \quad (13)$$
$$v_{dr} = R_r I_{dr} + p\psi_{dr} - \psi_{qr} p\beta \quad (14)$$
$$v_{qr} = R_r I_{qr} + p\psi_{qr} + \psi_{dr} p\beta \quad (15)$$

The flux linkages of the IM are represented by the symbol $\psi$ and are calculated using the following formulas:

$$\psi_{ds} = (L_{ls} + L_m)I_{ds} + L_m I_{dr} \quad (16)$$
$$\psi_{dr} = (L_{ls} + L_m)I_{dr} + L_m I_{ds} \quad (17)$$
$$\psi_{qs} = (L_{ls} + L_m)I_{qs} + L_m I_{qr} \quad (18)$$
$$\psi_{qr} = (L_{ls} + L_m)I_{qr} + L_m I_{qs} \quad (19)$$

where $L_{ls}$ and $L_{lr}$ represent the leakage-inductance of stator circuit and rotor circuit, respectively. $L_m$ is the mutual inductance between the rotor and stator circuits. $R_s$ and $R_r$ are the stator and rotor resistance, respectively. The non-linearity in the electrical part of the IM is due to the speed voltage terms.

In addition to the equations above, the mechanical part of the IM is defined by a single differential equation [6]:

$$p\omega_r = \frac{(T_e - T_L - D\omega_r)}{J} \quad (20)$$

where

$$T_e = \frac{3}{4} L_m poles (I_{dr} I_{qs} - I_{qr} I_{ds}) \quad (21)$$

and $T_e$ is the electrical torque of the IM in N.m and $J$ is the motor net inertia in kg.m$^2$. *poles* is the number of poles in the induction motor. The load torque ($T_L$) is generally described with a polynomial function of rotor speed.

The aforementioned equations map the transient behavior of a balanced three-phase squirrel cage IM into mathematical form. The mathematical model is then converted into an equivalent circuit using methods described in Section II and as shown here in Figure 4.

An extra equation is added to incorporate the zero sequence terms for the case of unbalance voltages at the motor terminals. If the motor were to have negative torque it would have to be separately calculated and added to (20).

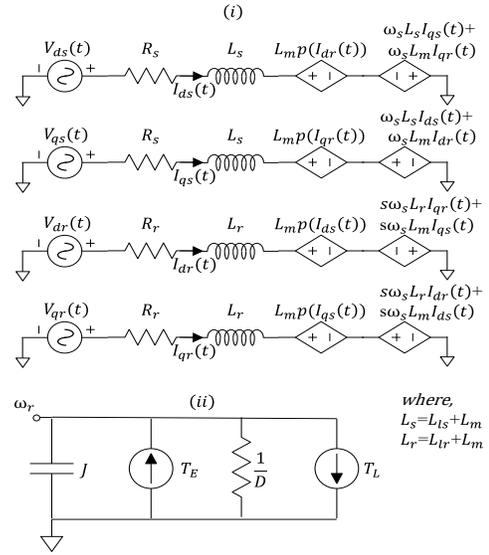

Figure 4: *Equivalent circuit for 3-phase induction motor: (i) Electrical circuit; and (ii) Mechanical Circuit*

V. VALIDATION

We now introduce SUGAR (Simulation with Unified Grid Analyses and Renewables), a grid simulation prototype tool that is designed to perform unified power system analysis. SUGAR and the IM model are validated by simulating the transient behavior of the IM during motor start-up and comparing the simulated results against those produced by SimPowerSystems© (SPS) module in Matlab. A 20 hp, 460 volts three-phase single squirrel cage IM model is used for the validation. The motor data is given in Table 1.

TABLE 1: 3-PHASE SQUIRREL CAGE INDUCTION MOTOR PARAMETERS

| $V_{LL}$ (Volts) | f (Hz) | $R_s$ (Ω) | $R_r$ (Ω) | $L_{ls}$ and $L_{lr}$ (mH) |
|---|---|---|---|---|
| 460 | 0.2761 | 0.2761 | 0.1645 | 2.191 |
| $L_m$ (mH) | poles | J (kg.m$^2$) | D (N.m.s) | $T_L$ (N.m) |
| 76.14 | 2 | 0.1 | 0.01771 | 10 |

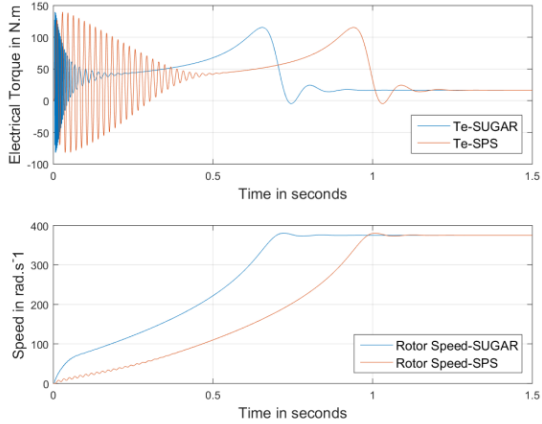

Figure 5: *Electrical Torque and Rotor Speed comparison between SimPowerSystems and SUGAR*

Figure 5 shows the response of IM's critical parameters during motor start-up. The evolution of motor state variables over time exhibit similar form and shape when simulated with both SPS and SUGAR. However, discrepancies are present in the magnitude of the results between the two. These discrepancies in the results are due to the approximation methods used by SPS in solving non-linear ordinary differential equations. SPS is documented to use a predictor-corrector numerical integration approach, which is extremely efficient, but does not create the Jacobian matrix of the system and does not use N-R for each time-step until convergence is reached. This can result in some loss of accuracy, as validated in Figure 6, which shows a plot of the electric torque of the motor using the SUGAR solver with the N-R iterations restricted to one per time step. With the SUGAR solver configured in this manner, it is equivalent to a predictor-corrector numerical integration approximation, and the results are identical to those produced by SPS.

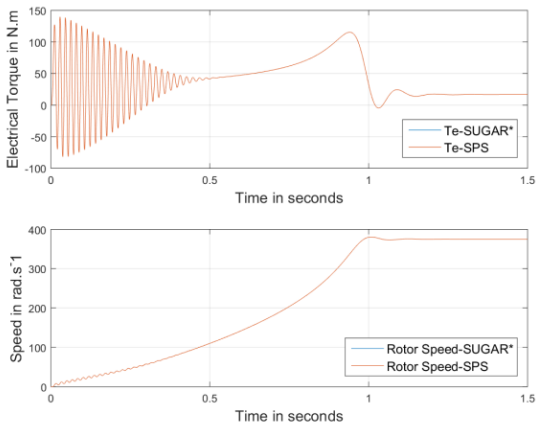

Figure 6: *Electrical Torque and Rotor Speed comparison between SimPowerSystems and SUGAR with SUGAR restricted to maximum of one N-R iteration.*

## VI. RESULTS AND DISCUSSION

Next, we validate the steady state solver of SUGAR for the case of a 3-phase IM model. For this example, the steady state solver solves the IM model in frequency domain with the frequency of the system set to the source frequency. The results from the steady state solver are then compared against the one obtained from the transient solver for the same IM. The transient analysis is run from t=0 to an approximate steady state condition at t=1.5 seconds. The comparison of the two simulations is presented in Table 2. The results are a perfect match to at least three significant digits.

TABLE 2: COMPARISON OF STEADY STATE AND TRANSIENT ANALYSIS IN SUGAR

| Parameter | Unit | Steady State | Transient @ t=1.5 sec |
|---|---|---|---|
| Rotor Speed | rad.s$^{-1}$ | 375.01 | 375.01 |
| Electric Torque | N.m | 16.64 | 16.64 |
| Stator direct-axis current | Amps | -11.36 | -11.36 |
| Stator quadrature-axis current | Amps | 13.09 | 13.09 |
| Rotor direct-axis current | Amps | 11.56 | 11.56 |
| Rotor quadrature-axis current | Amps | -0.49 | -0.49 |

## VII. CONCLUSION AND FUTURE WORK

Physics based models that best emulate the real physical state of the power grid are difficult and at times impossible to incorporate in traditional power flow algorithms. In this paper, circuit domain methods using equivalent circuit formulation with voltage and current as state variables have shown to incorporate any physics based model, without loss of generality. Furthermore, the paper has demonstrated that the use of physics based models along with equivalent circuit formulation facilitates unification of power system analyses, which provides consistent results between the steady state and transient power system analyses. Our prototype simulation tool SUGAR (Simulation with Unified Grid Analyses and Renewables) was demonstrated for an induction motor, but could be applied to any physics-based models in a similar manner.

As future work, we propose to explore alternative models for aggregated load and generation models (PQ/PV) that can best emulate real measured data from the grid. We also intend to study harmonic content in the power grid using circuit based harmonic balance methods. Importantly, we intend to use these harmonic balance methods to report consistent results between the transient and the steady state analyses for power systems that have a high percentage of nonlinear loads with substantial harmonic content.